\documentclass[conference]{IEEEtran}

\usepackage{cite}
\usepackage{amsmath,amssymb,amsfonts}
\usepackage{algorithmic}
\usepackage{graphicx}
\usepackage{textcomp}
\usepackage{xcolor}
\usepackage[hyphens]{url}

\def\BibTeX{{\rm B\kern-.05em{\sc i\kern-.025em b}\kern-.08em
    T\kern-.1667em\lower.7ex\hbox{E}\kern-.125emX}}

\begin{document}

% Ensure letter paper
\pdfpagewidth=8.5in
\pdfpageheight=11in

%%%%%%%%%%%%%%%%%%%%%%%%%%%%%%%%%%%%%%%%%%%%%%%%%%%%%%%%%%%%%%%%%%%%%%%%%%%%%%
% Misc. macros

\newcommand{\RISCV}{\mbox{RISC-V}}
\newcommand{\gemFiveCHERI}{\mbox{gem5-CHERI}}

\newcommand{\refsec}[1]{Sec.~\ref{sec:#1}}
\newcommand{\reffig}[1]{Fig.~\ref{fig:#1}}
\newcommand{\reftab}[1]{Tab.~\ref{tab:#1}}
\newcommand{\refalg}[1]{Alg.~\ref{alg:#1}}
\newcommand{\refdef}[1]{Def.~\ref{def:#1}}
\newcommand{\refeq}[1]{Eq.~(\ref{eq:#1})}
%%%%%%%%%%%%%%%%%%%%%%%%%%%%%%%%%%%%%%%%%%%%%%%%%%%%%%%%%%%%%%%%%%%%%%%%%%%%%%

\pagenumbering{arabic}

%%%%%%%%%%%---SETME-----%%%%%%%%%%%%%
\title{TechReport: Implementing CHERI in gem5}
\author{
	\IEEEauthorblockN{%
		Philipp Schmitz\IEEEauthorrefmark{1}\IEEEauthorrefmark{2},
		Vishwas Vasuki Gautam\IEEEauthorrefmark{1}\IEEEauthorrefmark{2},
		Tobias Jauch\IEEEauthorrefmark{2},
		Anna Lena Duque Ant\'{o}n\IEEEauthorrefmark{2},\\
		Johannes M\"{u}ller\IEEEauthorrefmark{2},
		Dominik Stoffel\IEEEauthorrefmark{2},
		Wolfgang Kunz\IEEEauthorrefmark{2}
	}
	\IEEEauthorblockA{
		\vspace{1ex}
		\IEEEauthorrefmark{2}RPTU University Kaiserslautern-Landau, Germany\\
		\IEEEauthorrefmark{1}Both authors contributed equally to this research.
	}
}
%%%%%%%%%%%%%%%%%%%%%%%%%%%%%%%%%%%%

\maketitle
\begin{abstract}
Despite countless mitigation efforts, memory safety issues still prevail.
The recently discovered security issues in PHP underline this fact.
One of the most promising approaches to solving this problem is the concept of Capability Hardware Enhanced RISC Instructions (CHERI).
There is a well-developed ecosystem of ISA simulators, compilers, operating systems, and hardware designs that support CHERI. 
\par
However, the lack of a simulator modeling microarchitectural detail hinders efficient and fast design space exploration for CHERI-enhanced  designs means the performance impact of CHERI remains understudied.
Therefore, we present \gemFiveCHERI: a CHERI-enhanced adaptation of the gem5 simulator that can boot the CheriBSD operating system and run custom workloads. 
The simulator is available open-source on GitHub.
\end{abstract}

\section{Introduction}
Memory safety is one of the oldest concerns in computer security~\cite{2013-SzekeresPayer.etal}.
Still, issues like out-of-bounds reads and writes continue to rank high in the top 10 of the MITRE CWE most dangerous software weaknesses in 2025~\cite{web-mitre-top25}. 
Widely used programming languages like C/C++ are not memory-safe as they allow for manual pointer modification and dereferencing which can cause memory leaks or corruption.
\par
This enables attackers to alter the control flow or behavior of the program, for example with the help of buffer overflows and code injection~\cite{2000-CowanWagle.etal}.
Defenses like Address Space Layout Randomization or data execution prevention impede memory corruption exploits but do not combat the underlying issue~\cite{web-pax-aslr}.
In addition, these defenses can be circumvented by other attack mechanisms, such as return-oriented programming~\cite{2012-RoemerBuchanan.etal}. New vulnerabilities related to memory safety continue to emerge. 
\par
In October 2026, the security issues in PHP related to CVE-2026-109618 and CVE-2026-109619 globally caused service disruptions~\cite{web-NYT-PHPissue}. 
Both are related to the known weaknesses \emph{Improperly Controlled Sequential Memory Allocation}, \emph{Use After Free}, \emph{Improper Handling of Exceptional Conditions}, and \emph{Access of Uninitialized Pointer}.
This highlights the need for holistic memory safety approaches, especially since rapidly increasing AI capabilities make it easier to detect and exploit such vulnerabilities. 
\par
One approach to tackle this issue is to use modern type and memory-safe languages like Rust~\cite{2014-MatsakisKlock}.
However, it is not possible to replace all C/C++ legacy code in the near future with Rust. For some code, such as certain low-level embedded software, a replacement can be technically infeasible~\cite{2024-SharmaSharma.etal}.
\par
A possible solution to this problem is to enforce memory safety directly in the hardware.
In recent years, the concept of \emph{Capability Hardware Enhanced RISC Instructions (CHERI)}~\cite{2014-WoodruffWatson.etal} is gaining traction as one of the most promising and holistic approaches to architectural memory safety.
CHERI's main concept are so called \emph{Capabilities} that extend ordinary pointers by appending meta data encoding the bounds and permissions plus an additional tag bit, at the cost of effectively doubling the original size.
CHERI extends customary \emph{Instruction Set Architectures} (ISAs) such as ARM and \RISCV{} with additional instructions to handle and manipulate these capabilities.

\section{Scope of this work}
The main aspects of this work are summarized as follows:
\begin{itemize}
    \item We extend the gem5 computer architecture simulator to support the CHERI-\RISCV{} v9 ISA. The extension comprises capability registers, CHERI support in the decoder, a tag-aware memory subsystem, and the enforcement of the CHERI bounds and permission checks. The simulator is available open-source\footnote{\url{https://github.com/RPTU-EIS/gem5-CHERI/}}. 
    \item Our simulator can boot the CheriBSD operating system in full system mode and run custom user-defined workloads. This enables fast DSE for CHERI workloads. 
    Additionally, CHERI-specific microarchitectural components can be prototyped and profiled.
\end{itemize}

\section{\gemFiveCHERI}
\label{sec:methodology}
\begin{figure*}[t]
  \centering
  \includegraphics[width=\textwidth]{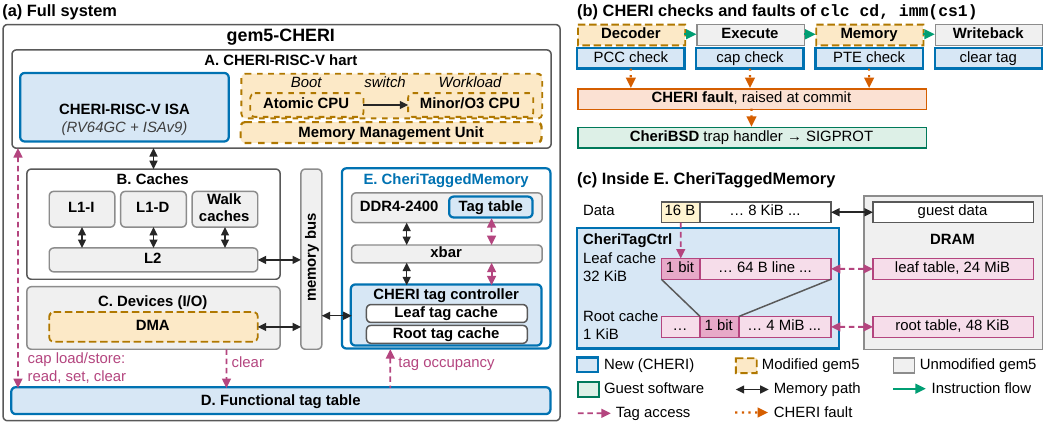}
\caption{(a) Overview of \gemFiveCHERI, (b) CHERI checks and faults in each stage of the instruction flow $clc$ $cd,$ $imm(cs1)$, and (c) Inside E. \emph{CheriTaggedMemory} with a two-level tag table  }
  \label{fig:overview}
\end{figure*}

Our goal is to enable the full system microarchitectural evaluation of CHERI-\RISCV{} single-core designs.
We need a platform that is both capable of booting an operating system such as CheriBSD~\cite{web-cheri-bsd} and running user workloads. 
Additionally, it must model microarchitectural detail and be easily configurable for design space exploration.
\par
We build \gemFiveCHERI{} on the baseline of the well-established gem5 computer architecture simulator~\cite{2011-BinkertBeckmann.etal, 2020-LowePowerAhmad.etal}.
The open-source community-supported project provides a Python-based scripting interface that makes the simulator dynamically configurable.
It provides well-tested standard components including out-of-order and in-order processor cores, caches, a DRAM model, I/O devices and more.
The components are designed as parameterized models, which makes it easy to explore different system configurations, e.g., evaluate the performance impact of different cache sizes.
More information on the gem5 simulator can be found in~\cite{2011-BinkertBeckmann.etal, 2020-LowePowerAhmad.etal}.
We base \gemFiveCHERI{} on the latest CHERI-\RISCV{} ISA v9~\cite{UCAM-CL-TR-987}.
\par
Integrating CHERI into a CPU design involves considerable microarchitectural adaptations to support the capability-based features.
These changes affect almost every part of a standard \RISCV{} implementation and change an instruction's latency along its way through the pipeline.
It is important for designers to be able to evaluate the impact of specific design decisions on the system performance quickly and easily.
\par
The main goals of our implementation are as follows:
\begin{itemize}
    \item \textbf{Full-system capability:} The simulator can boot a CheriBSD kernel and can run userspace workloads in pure-capability, hybrid and plain RV64 modes. 
    This ensures comparability between the different modes and enables realistic benchmarking and workload profiling.

    \item \textbf{Re-use of existing gem5 infrastructure:} The added CHERI support is integrated into gem5's existing CPU models, components and simulation infrastructure. 
    This simplifies architectural modifications, design-space exploration, and trustworthy performance metrics.  
\end{itemize}

\reffig{overview}(a) shows an overview of our \gemFiveCHERI{} implementation. 
The implemented features can be divided into three main categories. 
The blue blocks denote new CHERI components that are implemented from scratch. 
The dashed yellow blocks are components that already exist in gem5 but needed to be modified for supporting the CHERI specifications. 
Lastly, the grey blocks are components that remain unmodified. 
We implement CHERI as a separate ISA, derived from the RISC-V RV64GC ISA implementation of gem5~\cite{2017-RoelkeStan}. 
\par
The first modified component is the single CHERI-RISC-V hart (A) that implements the CHERI ISAv9. 
First, we created a 129 bit capability register container that is used for the register file and the 14 capability-extended control and status registers (CSRs).
We implemented a \emph{merged} register file, i.e., the integer register $x_n$ is the address field of the 129 bit capability register $c_n$.
All integer instructions modifying $x_n$ overwrite bits 128 to 64 of $c_n$ with zeros.
One important modification is the program counter capability (PCC).
Besides checking the allowed bounds during the instruction fetch stage, the PCC also encodes the current execution mode, i.e., \emph{capability} or \emph{integer}.
Depending on this mode some instructions are allowed to be executed or decoded differently, e.g., \texttt{AUIPC} becomes \texttt{AUIPCC}. 
The decode unit is extended to reflect this behavior and to support the compressed and uncompressed CHERI-specific instructions for capability management and accesses required to run custom workloads under CheriBSD.
\par
The memory management unit (MMU) translates the virtual addresses used by software into physical addresses and enforces the CheriBSD operating system's page permissions. 
In \gemFiveCHERI, we extend the paging control to capabilities as specified in the ISAv9. 
The reserved bits of the page table entry (PTE) format Sv39 are utilized to let the OS control loads and stores of capabilities. 
\par
The core part of the CHERI ISA are the formally verified memory safety guarantees~\cite{2020-NienhuisJoannou.etal, 2025-DuequeAntonMueller.etal} that are enforced by permission and bounds checks. 
In \gemFiveCHERI, we implement each check as an independent function performing the architectural checks on memory accesses, jumps, and instruction fetch.
We perform these checks in the priority order given in the Sail CHERI-RISC-V reference model~\cite{web-cheri-sail}, i.e., a capability must be tagged, unsealed, carry the required permission, and the access must be within the encoded bounds.
A failed check raises a CHERI exception that records which register and which check failed. 
Additionally, monotonicity is ensured by clearing the tag whenever a derived capability would exceed the original capability's rights.
\begin{figure}[t]\centering
  \includegraphics[width=\columnwidth]{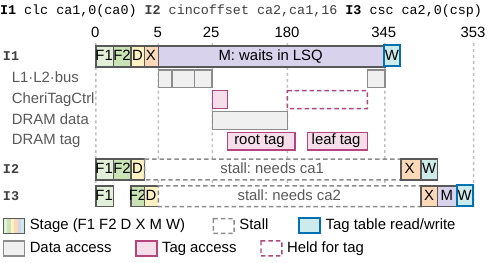}
  \caption{Example execution trace for a CHERI instruction sequence.}
  \label{fig:tagpipe}
\end{figure}

\reffig{overview}(b) shows where these checks happen at the example of the capability load  \texttt{clc~cd,~imm(cs1)}. 
The decoder checks that the PCC allows the instruction to be fetched. 
The execute stage computes the effective address from \texttt{cs1} and checks the tag, seal, permissions and bounds (i.e. cap check) of \texttt{cs1}. 
If a check fails, no request is sent to memory. 
During address translation, the MMU checks the capability permissions in the PTE.
If the page is not permitted to hold capabilities, the load completes but the tag
of the loaded value is cleared. 
If the page's capability generation is stale, the MMU raises a capability page fault,
which CheriBSD uses as its revocation load barrier. 
A faulting instruction carries its fault to the commit stage, meaning a squashed wrong-path instruction never traps. 
A CHERI exception then enters the CheriBSD trap handler, which sends a \texttt{SIGPROT} signal to the process that caused the fault.
\par
Our \gemFiveCHERI{} simulator supports all standard gem5 CPU models. All models are modified to replace the standard integer register file with the capability register class and the dependencies are tracked through the standard scoreboard and renaming stage.
In the o3 CPU model, we modified store-to-load forwarding to make it capability- and tag-aware and implemented a new type of memory request for the \texttt{cloadtags} instruction. 
\par
The cache subsystem (B in \reffig{overview}(a)) and I/O devices (C) are mostly left unmodified with the exception of direct memory access (DMA) components.
Every DMA write must clear the corresponding tags for memory regions it overwrites for enforcing safety. 
This ensures that a device does not turn an arbitrary write into a valid capability. 
\par
One major challenge in adopting the CHERI ISA is the management of the additional tag bit since standard addressing and accesses do not support 129 bit granules.
Therefore, we split the implementation into two parts: the functional tag table (FTT) (D) and the \emph{CheriTaggedMemory} (E).
\par
The FTT is a software construct which holds the only functional copy of every tag value. 
CHERI enforces a tag bit for every 16 byte physical address chunk.
The FTT is a hash map keyed by a chunk's address and only stores tagged chunks. 
Non-hashed values have the tag 0. 
All instructions access the FTT during the write-back stage, ensuring functional correctness. Lastly, as the FTT is a purely functional implementation, accessing it takes zero simulated cycles.
This choice vastly simplifies the implementation effort as it allows us to keep the caches unmodified.
However, we still want to faithfully model the tag-related timing. 
\par
Therefore, we introduce \emph{CheriTaggedMemory}.
We base our implementation on the ``Efficient tagged memory’’ from~\cite{2017-JoannouWoodruff.etal} by adding a small tag cache and with its own controller.
The controller manages a two-level tag table (see \reffig{overview}(c)) that is located in a reserved region at the top of physical memory.
\par
The lower level is the leaf tag memory.
One leaf bit holds the tag of one 16\,B granule, i.e., one capability.
Thus, one 64\,B leaf tag line covers the tags for 8\,KiB of data.
In total, the leaf memory occupies around 0.78\,\% of the protected DRAM memory, e.g., for 8 GiB of DRAM 64 MiB of leaf memory are required. 
\par
The root tag memory can help accelerate tag reads for memory regions.
A root bit is set if any tag in its leaf line is set, so a 64\,B root line covers 512 leaf lines, or 4\,MiB of data.
\par
Overall, both the tables round up to a 0.85\,\%  reservation of memory that the device tree hides from the CheriBSD kernel.
In our implementation, the controller performs DRAM requests to model the traffic and timing but the results are never propagated. Instead, the values are taken from the FTT.
\par
We implement a parametrized cache inside the controller to hold recently used table lines.
It can be disabled, configured in \emph{flat} mode, i.e., only use leaf tag memory, or,
by default, configured as a \emph{hierarchical} 32\,KiB 8-way leaf cache and a 1\,KiB 4-way root cache.
In hierarchical mode, both caches are independent and each reads its own table. The root level only decides whether a leaf line has to be read. If the root bit is 0, the leaf line is all zeros and the read is skipped.
\par
Hits in the L1 and L2 caches access tags in zero simulated time, faithfully modeling real hardware timing for CHERI-adapted caches.
Therefore, only requests that do not get served by the L2 cache reach the \emph{CheriTaggedMemory} controller that issues a tag request to model memory traffic.
Note that all requests to DRAM, independently of \emph{integer} or \emph{capability} access, create tag traffic since every 16 B chunk of data comes with a corresponding tag bit.
\par
On a leaf cache hit the tags are ready at no extra cost, since the lookup is shadowed by the DRAM data access. 
On a leaf cache miss, a tag-table read is issued two controller cycles after the lookup. Reads that go to the same line are merged by the controller. 
\par
In hierarchical mode, a root miss is served first, i.e., the root line is read from DRAM, and only when it returns as set, the controller checks the leaf bit.
A data reply that arrives before its tags is held in the controller until they arrive. 
This hold is the only latency tags add directly to a load, although tag traffic can also delay other requests through DRAM contention.
\par
Stores write their tags to the FTT in the writeback stage and issue a DRAM write request for the tag when the tag cache line is evicted. 
\par
Fig.~\ref{fig:tagpipe} shows an example execution trace. 
The capability load \texttt{I1} misses in the L1 and L2 and reaches the memory controller. 
Its data line is read from DRAM while the controller, having missed in both tag caches, reads the root line. 
Both DRAM requests return shortly after each other. 
The controller then reads the leaf line, and the data reply is held until that read returns one full DRAM access later. 
\texttt{I2} and \texttt{I3} depend on \texttt{I1} and stall. \texttt{I2} is a CHERI arithmetic instruction and does not involve a  memory access. 
The capability store \texttt{I3} hits in the L1 and sets its tag in the functional table without involving the controller. 

\section{Validation}
After implementation we validated that our implementation behaves according to the functional specification.
We run two functional test suites both on our \gemFiveCHERI{} simulator and on CHERI-QEMU~\cite{web-cheri-qemu} and compare the result of every test. 
\par
The first validation is based on CheriBSD's own regression test suite \emph{cheribsdtest}.
The tests check basic ISA-level behavior, e.g., capability bounds and permissions, sealing and sentries, capability loads and stores and their tags. 
Many tests deliberately cause a capability fault and make sure that the kernel delivers the expected \texttt{SIGPROT} signal with the expected cause. 
\par
We run the \emph{default} statically linked build of the test suite on QEMU, the AtomicCPU, the MinorCPU and the out-of-order (o3) CPU model and achieve matching results.
All 313 tests yield the expected \emph{pass} or \emph{fail} result.
\par
Second, we run the more comprehensive security benchmark suite Juliet 1.3~\cite{web-juliet} from the US National Institute of Standards and Technology (NIST).
Each test targets one class from the Common Weakness Enumeration (CWE) list, such as \emph{Stack Overflow} or \emph{Buffer Under-read}. 
We, again, achieve the same results as CHERI-QEMU for the 25,180 test cases from the Juliet test suite.
\par
Together, both experiments validate our implementation of the CHERI memory safety features.

\section{Conclusion}
With \gemFiveCHERI{}, we present the first CHERI-aware microarchitectural simulator, allowing for fast and efficient design space exploration.
We extend gem5 with the CHERI-RISC-V ISAv9, including the functional and faulting behavior of all instructions as well as the tag management in memory.
We decouple the tag functionality from the timing model in order to be able to investigate the introduced overheads individually.
In future work, \gemFiveCHERI{} could be extended to support multi-cores, i.e., modeling cache coherence for tags. 

\section*{Acknowledgement}
The reported research was supported by Agentur für Innovation in Cybersicherheit GmbH under grant PROTECT (EVIT program).

%%%%%%%%%%%%%%%%%%%%%%%%%%%%%%%%%%%%%%%%%%%%%%%%%%%%%%%%%%%%%%%%%%%%%%%%%%%%%%%%
% References
%
\bibliographystyle{IEEEtran}
\bibliography{refs3}
%%%%%%%%%%%%%%%%%%%%%%%%%%%%%%%%%%%%%%%%%%%%%%%%%%%%%%%%%%%%%%%%%%%%%%%%%%%%%%%%

%%%%%%%%%%%%%%%%%%%%%%%%%%%%%%%%%%%%%%%%%%%%%%%%%%%%%%%%%%%%%%%%%%%%%%%%%%%%%%%%
% That's all folks
%
\end{document}